\documentclass[12pt]{article}

\usepackage[dvips, letterpaper,hmargin=1in,vmargin=1in]{geometry}

\usepackage{graphicx,amsmath,amsfonts,amssymb,aas_macros}

\usepackage{titlesec}
\titleformat{\section}{\large\bfseries}{\thesection}{1em}{}

\renewcommand{\sec}{\ensuremath{\mathrm{s}}}

\newcommand{\kg}{\ensuremath{\mathrm{kg}}}

\newcommand{\fm}{\ensuremath{\mathrm{fm}}}
\newcommand{\cm}{\ensuremath{\mathrm{cm}}}
\newcommand{\km}{\ensuremath{\mathrm{km}}}

\newcommand{\eV}{\ensuremath{\mathrm{eV}}}
\newcommand{\keV}{\ensuremath{\mathrm{keV}}}
\newcommand{\MeV}{\ensuremath{\mathrm{MeV}}}
\newcommand{\GeV}{\ensuremath{\mathrm{GeV}}}

\newcommand{\bet}{\ensuremath{{}^8\mathrm{B}}}
\newcommand{\hef}{\ensuremath{{}^4\mathrm{He}}}

\newcommand{\keVee}{\ensuremath {\mathrm{keVee}}}

\newcommand{\Er}{\ensuremath {E_R}}
\newcommand{\Ev}{\ensuremath {E_\mathrm{v}}}
\newcommand{\cpd}{\ensuremath {\mathrm{cpd}}}
\newcommand{\days}{\ensuremath {\mathrm{days}}}
\newcommand{\ton}{\ensuremath {\mathrm{ton}}}
\renewcommand{\day}{\ensuremath {\mathrm{day}}}

\newcommand{\dmb}{\ensuremath {\Delta m_b^2}}
\newcommand{\nub}{\ensuremath {\nu_{b}}}
\newcommand{\nubb}{\ensuremath {\overline\nu_{b}}}
\newcommand{\DM}{\ensuremath {\mathrm{DM}}}

\newcommand{\Neff}{\ensuremath {\mathcal{N}_{\mathrm{eff}}}}
\newcommand{\N}{\ensuremath {\mathcal{N}}}

\newcommand{\Enu}{\ensuremath {E_\nu}}

\begin{document}

\begin{titlepage}

\setcounter{page}{1}

\vspace*{0.2in}

\begin{center}

\hspace*{-0.6cm}\parbox{17.5cm}{\Large \bf \begin{center}
Elastic scattering signals of solar neutrinos with enhanced 
baryonic currents
\end{center}}

\vspace*{0.5cm}
\normalsize

\vspace*{0.5cm}
\normalsize

{\bf Maxim Pospelov$^{\,(a,b)}$ and Josef Pradler$^{\,(a)}$}

\smallskip
\medskip

$^{\,(a)}${\it Perimeter Institute for Theoretical Physics, Waterloo,
ON, N2L 2Y5, Canada}

$^{\,(b)}${\it Department of Physics and Astronomy, University of Victoria, \\
  Victoria, BC, V8P 1A1 Canada}

\smallskip
\end{center}
\vskip0.2in

\centerline{\large\bf Abstract}

The coupling of the baryonic current to new neutrino states \nub\ with
strength in excess of the weak interactions is a viable extension of
the Standard Model.
We analyze the signature of \nub\ appearance in the solar neutrino
flux that gives rise to an elastic scattering signal in dark matter
direct detection and in solar neutrino experiments.
This paper lays out an  in-depth study of \nub\ detection prospects for
current and future underground rare event searches.
We scrutinize the model as a possible explanation for the reported
anomalies from DAMA, CoGeNT, and CRESST-II and confront it with
constraints from other null experiments.

\vfil
    
\end{titlepage}

\section{Introduction}

The phenomenon of neutrino mass-mixing and thereby induced
lepton-flavor oscillations constitute the only conclusively detected
deviation from the Standard Model (SM) to date. Though the neutrino
sector is certainly the most elusive part of the SM, a tremendous and
sometimes painstaking experimental effort has firmly established the
patterns of mass splitting and mixing for the three SM neutrinos,
$\nu_e,\,\nu_{\mu},\,\nu_{\tau}$.  The multitude of data is consistent
with neutrino flavor eigenstates being the linear combination of---at
least---three massive ones $\nu_1,\,\nu_2,\,\nu_3$, with mass-squared
differences $\Delta m_{21}^2 = \mathcal{O}(10^{-5})\,\eV^2$ and $
\Delta m_{31}^2 \sim \Delta m_{32}^2 = \pm\mathcal{O}(10^{-3})\,\eV^2$
and mixing angles with magnitudes $\sin^22\theta_{13}\sim 0.1$,
$\theta_{23}\sim \pi/4$, $\tan^2\theta_{12} \sim 1/2
$~\cite{Nakamura:2010zzi,An:2012eh}.  Neutrinos are predominantly detected via
their charged current (CC) interaction on matter, and elastic
scattering (ES) on electrons.  The only pure neutral current (NC)
process for solar neutrinos with participation of all active neutrino
flavors observed remains the deuteron breakup reaction $d+\nu_x\to p
+n +\nu_x$ in the SNO experiment~\cite{Ahmad:2002jz,Aharmim:2011vm}.

On the theoretical side, the interest to the elastic scattering of
neutrinos on nuclei dates back to Drukier and
Stodolsky~\cite{PhysRevD.30.2295}, who outlined their vision of a true
neutrino observatory: neutrinos with MeV-scale energies as they emerge
from the interior of the sun, from nuclear reactors, from spallation
beam experiments, or from supernovae explosions can scatter
elastically and with $N^2$ enhancement in the cross section on
keV-scale recoiling nuclei of $N$~neutrons~\cite{Freedman:1977xn}.
Even though the energy transfer to the target nucleus is small, the
idea was that its effect in a detector at cryogenic temperatures may
be macroscopic given that the specific heat of the material can be
minute. The underlying principle of NC neutrino scattering on nuclei
has found its proliferation in Dark Matter (DM) searches for weakly
interacting massive particles (WIMPs)~\cite{PhysRevD.31.3059}.

Direct detection experiments seek for evidence of DM via its elastic
scattering on various targets such as crystals made from germanium or
liquefied nobles gases such as xenon.
No univocal evidence for DM has yet been found but upper limits as low
as $7\times 10^{-45}\,\cm^2$~\cite{Aprile:2011hi} in the WIMP-nucleon
cross section have been inferred for the optimal range of recoil
energies.
These experiments have reached a level in sensitivity such that
neutrino coherent scattering on nuclei is being discussed as potential
background for future ton-scale experiments; see
\textit{e.g.}~\cite{Monroe:2007xp,Strigari:2009bq}.
The latter process has a cross section $\sigma_{\nu}/N^2 \simeq
4\times 10^{-43} (E_{\nu}/10\,\MeV)^2\,\cm^2$ where $E_{\nu}$ is the
neutrino energy. Indeed, the average flux of neutrinos at earth,
dominated by the solar $pp$-chain of thermonuclear reactions,
$\Phi_{\nu}^{pp} \simeq 6\times 10^{10} \,
\cm^{-2}\sec^{-1}$~\cite{RevModPhys.60.297}, exceeds the flux expected
from WIMPs, $\Phi_{\DM} \sim 10^5 \left( 100\,\GeV/m_{\DM} \right)\,
\cm^{-2}\sec^{-1} $, by many orders of magnitude. The reason why
neutrinos are \textit{not} copiously detected via their NC interaction
in DM experiments lies in the soft recoil spectrum they induce, $ E_R
\leq {2 E_{\nu}^2}/{m_N} \simeq 2\,\keV \times \left( {100}/{A}
\right) \left( {E_{\nu}}/{10\,\MeV} \right)^2$, where $m_N$ and $A$
are the mass and atomic mass number of the target nucleus. Whereas
these experiments fall short in sensitivity to SM neutrinos, they may
nonetheless be powerful probes of an extended neutrino sector
\cite{Pospelov:2011ha}.

This paper surveys potential signatures of the existence of yet
another ``sterile-active'' neutrino state,~\nub, where active is
understood in the sense that \nub\ shall interact via a \textit{new}
neutral current interaction with baryons
(NCB)~\cite{Pospelov:2011ha}. This ``baryonic'' neutrino is also
``sterile'' in that it shares no SM NC or CC interactions. In
particular, this implies that \nub\ does not scatter elastically on
electrons.  Remarkably, \nub\ can then be coupled to baryons with a
strength $G_B$ which exceeds that of the weak interactions
substantially, $G_B/G_F = (10^2-10^3)$; $G_F$ is the Fermi constant.
This, in turn, implies the existence of a force mediator with the mass
much smaller than $m_{W}$.
The key observation made in \cite{Pospelov:2011ha} is that for
MeV-scale energies of \nub\ the ratio of elastic to inelastic cross
section with nuclei scales as~
\begin{align}
\label{eq:scaling}
  \frac{\sigma_{\nub N}(\mathrm{elastic})}{\sigma_{\nub
      N}(\mathrm{inelastic}) } \sim 10^8 \times \, \left( \frac{A}{100}
  \right)^2 \left( \frac{10\,\MeV}{E_{\nu}} \right)^4 \left(
    \frac{2\,\fm}{R_N} \right)^4 ,
\end{align}
where $R_N$ denotes a nuclear radius.  This tremendous enhancement
opens an exciting phenomenological window of opportunity: \nub\ states
can be searched for in the growing number of direct detection dark
matter experiments.  The less constrained choice of parameters for
this model is when the oscillation length for $\nu_{SM}\to \nu_b$
transition is long enough not to create any significant fluxes of
reactor/beam/atmospheric neutrinos, while for solar neutrinos one can
have a sizable fraction of $\nu_b$. In particular, solar \bet\
neutrinos have the best combination of large flux and high end-point
energy for producing an observable signal in rare event searches.
Relation (\ref{eq:scaling}) makes even small scale DM searches
competitive with dedicated large target-mass neutrino experiments in
their sensitivity to \nub\ that originate from $^8$B neutrinos.

Moreover, as proposed in~\cite{Pospelov:2011ha}, if the oscillation
length of \nub\ is on the order of the earth-sun distance, the signal
may be annually modulated in a non-trivial way, with a possibility of
a bigger flux in the summer, and with modulation amplitude larger than
naively expected.  In this paper we analyze this corner of the
parameter space in great detail, as it could offer a
$\nu_b$-scattering explanation to the DAMA
signal~\cite{Bernabei:2008yi,Bernabei:2010mq} while still being
compatible with other null-results from DM experiments; for a relevant
collider study in this context cf.~\cite{Friedland:2011za}. The
long-standing DAMA ``anomaly'' has very recently received some
additional impetus by the reports~\cite{Aalseth:2010vx,Aalseth:2011wp}
of the CoGeNT collaboration on 1) an unexpected rise in observed
events at low nuclear recoils and 2) an indication of annual
modulation, see
\textit{e.g.}~\cite{Aalseth:2011wp,Hooper:2011hd,Fox:2011px,HerreroGarcia:2011aa}.
Finally, the CRESST-II experiment has published results of its latest
run, which had some population of unexpected events on top of existing
backgrounds~\cite{Angloher:2011uu}.

The attempts to explain positive signals (DAMA) and possible hints on
non-zero signals (CoGeNT, CRESST) while staying consistent with null
results of other groups, are widespread in the WIMP literature, see
\textit{e.g}~\cite{Chang:2010yk,Feng:2011vu, Cline:2011zr,
  Schwetz:2011xm, Farina:2011pw, Frandsen:2011ts, Kopp:2011yr,
  Kelso:2011gd,Shoemaker:2011vi} and references therein. The models that fare best
feature $\sim 10$ GeV WIMP masses, although the overall consistency of
the "light WIMP" explanation for the "DM anomalies" remains
doubtful. In this paper, we provide an in-depth critical assessment of
$\nub$ models with regard to their potential to explain positive
signals, and make predictions for the upcoming experiments, paying
attention to those that could potentially differentiate between \nub\
and light dark matter recoils. We leave other phenomenological aspects
of this interesting, but to date a relatively poorly explored model to
subsequent work.

This paper is organized as follows: in the next section, the baryonic
neutrino model is reviewed. In Sec.~\ref{sec:dd} we cover the current
and future sensitivity of DM searches to \nub\ and study the potential
explanation of the DAMA, CoGeNT, and CRESST-II signals. In
Sec.~\ref{sec:ns} the elastic scattering of \nub\ in neutrino
experiments is considered, and in Sec.~\ref{sec:conclusions} we reach
our conclusions.

\section{Baryonic neutrinos}
\label{sec:nub}

In the baryonic neutrino model~\cite{Pospelov:2011ha}, the SM gauge
group is extended by an abelian factor U(1)$_B$. The new neutrino is a
left-chiral field $\nub = \frac{1}{2} (1-\gamma^5) \nu_b$ with charge
$q_b = \pm 1$ and gauge coupling $g_l>0$.
Leptons are neutral under U(1)$_B$ but all quarks $q=Q_L,\,u_R,\,d_R$
carry baryonic charge $1/3$ with gauge coupling $g_b > 0$.  The SM
Lagrangian is supplemented by the following terms
\begin{align}
  \mathcal{L}_{B} = \nubb \gamma^{\mu} (i \partial_{\mu} - g_l q_{b}
  V_{\mu}) \nub - \frac{1}{3} g_b \sum_q \bar q \gamma^{\mu}  q V_{\mu}
  -\frac{1}{4} V_{\mu\nu}V^{\mu\nu} + \frac{1}{2} m_V^2 V_{\mu}V^{\mu}
  + \mathcal{L}_{m} .
\end{align}
We have assumed that the new gauge field $V_{\mu}$ with field strength
$V_{\mu\nu}$ has acquired a mass~$m_V$ by the spontaneous breakdown of
the U(1)$_B$ symmetry with Higgs$_b$ VEV $\langle \phi_b \rangle =
v_b/\sqrt{2}$; the sum in the second term runs over all SM quarks
$q$. The part $\mathcal{L}_m$ is responsible for generating neutrino
masses and the mixing between SM neutrinos and the new state \nub. A
simple UV-completion of $\mathcal{L}_m$ is one where---once
electroweak symmetry and U(1)$_b$ are broken---new right-handed
neutrinos $\nu_R$ induce mass terms for SM neutrinos as well as for
\nub:
\begin{align} \label{eq:Lm}
  \mathcal{L}_m = \frac{1}{2} N_L^T  \mathcal{C}^{\dag} M N_L + \mathrm{h.c.}
  ,\quad N_L =
  \begin{pmatrix}
    \nu_L' \\ \nu_R'^C \\ \nub'
  \end{pmatrix},\quad
    M = 
\begin{pmatrix}
   0 & m_D^T  & 0\\
   m_D & m_R & v_b b \\
   0 & v_b b^T & 0
  \end{pmatrix} .
\end{align}
Here, $\nu'_L$ denote the three SM neutrinos with Dirac mass matrix
$m_D$, $\nu_R'^C = \mathcal{C} \overline{\nu_R'}^T$ are the charge
conjugate states of $\nu_R'$ with Majorana mass matrix $m_R$, and $b$
is a vector of Yukawa couplings generating mass for $\nu'_{b}$. 
In the simplest case the new right-handed neutrinos generate masses
for $\nu'_L$ and $ \nub'$ simultaneously. Introduction of a
right-handed partner for $\nub'$ will cancel a U(1)$_B^3$ triangle
anomaly and remaining gauge anomalies can be cured by the introduction
of a new family of heavy fermions with appropriate quantum
numbers~\cite{PhysRevLett.74.3122}. From the phenomenological
viewpoint the details of this will not be of importance for this work.

The mass matrix $M$ in~(\ref{eq:Lm}) is diagonalized by
$\mathcal{M}_{\mathrm{diag}} = (V_L^{\nu})^\dag M V_L^{\nu} $ where
$V_L^{\nu}$ is a unitary matrix, defining the mass eigenstates $N_L =
(V_L^{\nu})^{\dag} N'_L$.
Diagonalization of the charged lepton mass matrix by $3\times3$
unitary matrices $V_{L,R}^{l}$ with mass eigenstates $l_L =
(V_L^l)^{\dag} l'_L$, and $l_R = (V_R^l)^{\dag} l'_R$ where $l =
(e^{-}, \mu^{-},\tau^{-})$ then determines the mixing among the SM
active neutrino states.
Assuming that the eigenvalues of $m_R$ are much larger than any other
values in $M$, the seesaw mechanism is operative and one can integrate
out the heavy, right handed states. We are left with the $4\times4$
mixing matrix $U$ which connects ``flavor'' $\nu_{\alpha L}\,
(\alpha=e,\mu,\tau,b)$ and mass $n_{k L}\, (k=1,\,\dots,4)$
eigenstates:
\begin{align}
\label{eq:fltoma}
  n_{kL} = \sum_{\alpha } U^{*}_{k\alpha}
  \nu_{\alpha L}, 
\quad U = 
\begin{array}{cc}
  \hphantom{b} & \begin{matrix}\!\!\!\! 1 & 2 & 3 & 4 \end{matrix} \\
  \begin{matrix}e\\\mu\\\tau\\b\end{matrix}  & \!\!
  \begin{pmatrix} & & & \cdot\,\, \\  \multicolumn{3}{c}{U_{\mathrm{PMNS}}} & \cdot\,\,  
\\ & & &\cdot\,\,\\ \cdot\,\, & \cdot\,\, & \cdot\,\, & \cdot\, \, 
\end{pmatrix}\end{array} ,
\end{align}
where $U_{\mathrm{PMNS}}$ is the usual $3\times3$ mixing matrix among
the active flavors~\cite{Nakamura:2010zzi}.  The NCB current in the
respective interaction and mass eigenbasis reads
\begin{align}
  j_{NCB}^{\mu} = \nubb \gamma^{\mu} \nub  = \sum_{k,k'}\, U^{*}_{ 4 k}U_{4 k'} 
  \overline{n}_{kL}  \gamma^{\mu} n_{k'L}  .
\end{align}

\subsection{Neutrino oscillations and matter effects}
\label{sec:matter-effects}

Aiming at a scenario in which the baryonic neutrino is coupled
stronger than $G_F$, the question which immediately arises is whether
new matter effects are to play a role.
The index of refraction of \nub-propagation in matter is found by
computing the forward scattering amplitude of $\nub$ on matter,
described by the effective Lagrangian
\begin{align}
\label{eq:Leff}
  \mathcal{L}_{\mathrm{eff}} = - G_B j^{\mu}_{NCB} \sum_{N = n,p}
  \overline N \gamma_{\mu} N ,\qquad G_B = q_b \frac{g_b g_l }{
    m_V^2} = q_b \N G_F .
\end{align}
In the last equality we have introduced the parameter $\N>0$ to
measure $G_B$ in units of $G_F$ with the sign of the interaction
determined by the charge $q_b$ of \nub.
In an unpolarized medium, one obtains the following matter potential
\begin{align}
  V_{NCB} = \pm q_b \N G_F n_B \left( Y_N + 2Y_{\nu_b} \right), \qquad Y_f =
  \frac{n_{f} - n_{\overline f}}{ n_B},
\end{align}
where the plus sign is for \nub\ and the minus sign for \nubb;
$Y_f$ is the particle-antiparticle asymmetry, normalized to the number
density of baryons $n_B$. The first term in the first equation
describes the potential in ordinary matter while the second term is
the potential for $\stackrel{(-)}{\nu}_{\!\!bL}$ in a hypothetical sea
of baryonic neutrinos.

In ordinary matter, with mass fraction $X_p$ in form of bound or
unbound protons, the induced matter potentials (up to coherence
factors) compare as follows
\begin{align}
 V_{NCB} : V_{CC} : V_{NC} = q_b \N : \sqrt{2} X_p : -\sqrt{2} (1-X_p)/2 ,
\end{align}
where we have made use of the charge neutrality condition.
As is evident, for $\N\gg1$ baryonic neutrinos experience the
largest matter effect since $X_p$ is always of order unity. 

The large strength of the NCB interaction may suggest that the flavor
evolution in matter is dominated by $ V_{NCB}$. In a simplified
two-neutrino case the Schr\"odinger-like equation describing the
transition probabilities $P_{\alpha\to \beta}(x) = | \langle
\nu_{\beta} |\nu_{\alpha}(x) \rangle |^2 \equiv |\psi_{\alpha
  \beta}(x)|^2$ from an initial state $| \nu_{\alpha} (0)\rangle$ to
final state $|\nu_{\beta} \rangle$ can then be brought into the
following form
\begin{align}
  i \frac{d}{dx}
  \begin{pmatrix}
    \psi_{\alpha\alpha}\\ \psi_{\alpha b}
  \end{pmatrix} 
 \simeq
 \frac{1}{4\Enu}
 \begin{pmatrix}
   -\Delta m^2 \cos 2\theta-2\Enu V_{NCB} & \Delta m^2 \sin 2\theta \\
    \Delta m^2 \sin 2\theta  & \Delta m^2 \cos 2\theta + 2\Enu V_{NCB}
 \end{pmatrix}  
\begin{pmatrix}
    \psi_{\alpha\alpha}\\ \psi_{\alpha b}
  \end{pmatrix}  .
\end{align}
Here, $\Enu$ is the neutrino energy and $\theta = \theta_{k 4}$ and
$\Delta m^2 = \Delta m_{4k}^2$ with $k=1,2,3$ for $\alpha=e,\mu,\tau$
are the vacuum mixing angle and the mass squared difference between
the new massive state $\nu_4$ and SM neutrinos $\nu_{1,2,3}$,
respectively.  The matter induced mixing angle $\theta_M$ reads
\begin{align}
\label{eq:thetaM}
  \tan 2\theta_M = \frac{\tan 2\theta}{1+ 2\Enu V_{NCB} / (\Delta m^2 \cos
    2 \theta ) },
\end{align}
and, since $\mathrm{sign\,}(V_{NCB}) = q_b$, resonant flavor
transitions for $\nub$ could be possible for $q_b = +1$ and $\theta >
\pi/4$ or for $q_b = -1$ and $\theta < \pi/4$ (and vice verse for
\nubb.)  We note, however, that the efficiency of a transition has a
separate dependence on $\Delta m^2$, unrelated to
Eq. (\ref{eq:thetaM}), as the matter-induced oscillations seize to
occur in the limit of $\Delta m^2\to 0$.  From (\ref{eq:thetaM}) we
find the necessary condition for which NCB effects are least likely to
play a role,
\begin{align}
\label{eq:matter}
  \Delta m^2 \cos{2\theta} \ll 10^{-4}\,\eV^2 \times \left(
    \frac{E}{10\,\MeV} \right) \left( \frac{\N}{100} \right)
\left( \frac{\rho}{\mathrm{g}/\cm^3} \right)  .
\end{align}
In this work we focus on a parameter region which obeys this limit. A
discussion for larger values of $ \Delta m^2$ is beyond the scope of
this work.

\subsection{Solar \boldmath$\nu_b$ flux}

Let us consider a scenario in which the baseline of $\nub$ oscillation
$L_{\mathrm{osc}}$ is on the order of the earth-sun distance, $L_0 =
1\,\mathrm{AU} \simeq 1.5\times 10^8\,\km$. This ``just-so'' choice of
parameters suggests a canonical mass squared difference,
\begin{align}
\label{eq:Losc}
  \frac{L_{\mathrm{osc}}}{L_0} \simeq 0.5 \times \left(
    \frac{10^{-10}\,\eV^{2}}{\Delta m^2} \right) \left(
    \frac{E_{\nu}}{10\,\MeV} \right) .
\end{align}
``Flavor'' eigenstates $\nu_{\alpha L}$ $(\alpha = e,\mu,\tau,b)$ are
found from mass-eigenstates $n_{kL}$ by inversion
of~(\ref{eq:fltoma}), $ \nu_{\alpha L } = \sum_{k} U_{\alpha k} n_{k
  L}$, and their evolution is obtained by solving
\begin{align}
\label{eq:schroedinger}
  i \frac{d\Psi}{dx} = \mathcal{H} \Psi,\quad \mathcal{H} =
  \frac{1}{2E} \left( U \mathcal{M}_d^2 U^{\dag} + \mathcal{A} \right).
\end{align}
Here $\Psi$ is the vector of amplitudes for the flavor states,
$\Psi=(\psi_e,\psi_{\mu},\psi_{\tau}, \psi_b)$, $\mathcal{M}_d^2 =
\mathrm{diag}(m_1^2,m_2^2,m_3^2,m_4^2)$ is the diagonalized mass
matrix and the entries of $\mathcal{A} =
\mathrm{diag}(A_{CC}+A_{NC},A_{NC},A_{NC},A_{NCB})$ are related to the
induced matter potentials via $A_x = 2E V_x$. In general, the baryonic
neutrino flux at the Earth location is found upon numerical
integration of (\ref{eq:schroedinger}) from the production point $r_0$
of $\nu_e$ in the solar interior to earth at distance $L$ with initial
condition $\Psi(r_0) = (1,0,0,0)$, This could be a complex problem
when matter effects are involved, but fortunately not for the region
of the parameter space we are interested in.

With a few simplifying assumptions the appearance probability at earth
can be obtained analytically~\cite{Pospelov:2011ha}. We seek access to
the high energy end of the neutrino spectrum, $\Enu\gtrsim 10\,\MeV$,
because scatterings of $\nub$ will then more likely be picked up by a
detector.  The largest flux in combination with high endpoint energy
comes from the neutrino emission in the decay of \bet. With \hef\
being the most tightly bound light nucleus, hep neutrinos have the
highest endpoint in energy but come with a flux which is smaller by
three orders of magnitude. The \bet\ and hep respective fluxes and
endpoint energies are given by~\cite{2005ApJ...621L..85B},
\begin{align}
  \Phi_{\bet} = (5.69^{+0.173}_{-0.147})\times 10^6\,\cm^{-2}\,\sec^{-1},\quad  E_{\mathrm{max},
    \bet} = 16.36\,\MeV, \nonumber \\
  \Phi_{\mathrm{hep}} = (7.93\pm0.155 )\times 10^3\,\cm^{-2}\,\sec^{-1},\quad  E_{\mathrm{max},
    \mathrm{hep}} = 1.88\,\MeV . 
\label{eq:fluxes}
\end{align}
In the solution to the solar neutrino problem the MSW effect
dominates the flavor evolution of the highly energetic part of the
neutrino spectrum and neutrinos exit the sun mainly as
$\nu_2$. Therefore, we assume a preferential mixing of the new
neutrino to $\nu_2$, $ \theta_{24} \neq 0$, and neglect other mixings
of $\nu_4$ to the active states.
Choosing $q_b>0$, no resonant flavor transitions to $\nub$ inside the
sun appear and with a ballpark of $\Delta m^2$ suggested
in~(\ref{eq:Losc}) the standard solar MSW solution remains in place.

With these assumptions and using a tri-bimaximal mixing ansatz for the
active states the following $\nub$ appearance probability at earth for
the high energy part of the $\bet$ (and hep) fluxes has been obtained
in~\cite{Pospelov:2011ha},
\begin{align}
\label{eq:Peb}
  P_{eb}(L,\Enu) \simeq \sin^2(2\theta_b) \sin^2{\left[ \frac{\dmb
        L(t)}{4 \Enu} \right]} . 
\end{align}
It is assumed that mass mixings among active components are larger
than mixings with $\nub$ so that one can address the diagonalization
of the neutrino mass matrix sequentially. In this procedure, $\dmb$
and $\theta_b$ denote the associated effective mass-squared difference
and mixing angle between $\nu_2$ and $\nub$, respectively. The true
vacuum mass eigenstates are $\nu_I = \cos{\theta_b}\nu_2 +
\sin{\theta_b}\nu_b$ and $\nu_{II} = -\sin{\theta_b}\nu_2 +
\cos{\theta_b}\nu_b$ and a phase builds up between $\nu_2$ exit from
the sun and propagation to the detector at distance $L$,
\begin{align}
  L(t) = L_0 \left\{ 1-\epsilon \cos{\left[ \frac{2\pi(t-t_0)}{T}
      \right]} \right\} ,
\end{align}
where $\epsilon = 0.0167$ is the ellipticity of the earth's orbit; the
perihelion is reached at $t_0 \sim 3\,\mathrm{Jan}$.
In what follows, it will be convenient to introduce an effective
interaction parameter \Neff,
\begin{align}
  \label{eq:neff}
 \Neff^2 = \N^2 \, \sin^2(2\theta_b) / 2 .
\end{align}
In the limit of rapid oscillations this implies $P_{eb}G_B^2\to \Neff^2G_F^2$.

\section{Direct Detection}
\label{sec:dd}

In this section we provide a detailed investigation of current and
future direct detection experiments. From the
scaling~(\ref{eq:scaling}) we expect that elastic scattering off
nuclei in direct detection experiments constitutes one of the most
promising avenues in the search for a solar, long-baseline flux of
$\nub$ particles.

The spin-independent elastic recoil cross section on nuclei obtained
from~(\ref{eq:Leff}) essentially resembles the one from
neutrino-nucleus coherent scattering~\cite{PhysRevD.30.2295} with the
replacement $G_F^2(N/2)^2\to G_B^2 A^2$~\cite{Pospelov:2011ha},
\begin{align}
\label{eq:dsdcosth}
  \frac{d\sigma_{\mathrm{el}}}{d\cos\theta_{*}} = \frac{G_B^2}{4\pi}
  \,\Enu^2 A^2 (1+\cos\theta_{*}) .
\end{align}
Here $A$ is the atomic number of the nucleus and $\theta_{*}$ is the
scattering angle in the CM frame. Equation~(\ref{eq:dsdcosth}) can be
rewritten in terms of a recoil cross section,
\begin{align}
  \frac{d\sigma_{\mathrm{el}}}{d\Er} = \frac{G_B^2}{2\pi} \,A^2
  m_N  F^2(|\mathbf{q}|)\left[ 1 - \frac{(E_{\mathrm{min}})^2}{\Enu^2} \right] , \quad
  \Er = \frac{\Enu^2}{m_N} \left( 1 - \cos{\theta_{*}} \right),
\end{align}
where $E_{\mathrm{min}} = \sqrt{\Er m_N/2}$ is the minimum energy
required to produce a recoiling nucleus of mass $m_N$ and kinetic
energy $\Er$. Here we have now included the nuclear form factor
suppression $F^2(|\mathbf{q}|)$ for scatterings with three-momentum
transfer~$\mathbf{q}$, and in our numerical evaluations we use the
Helm form factor parametrization~\cite{Helm:1956zz} with the nuclear
skin thickness of 0.9\,\fm.

For the sake of comparison to the simplest case of spin-independent
scattering of DM on nuclei, we can evaluate the total elastic
scattering cross section of \nub\ on nuclei (at zero momentum
transfer),
\begin{align}
\label{eq:sigTotal}
\sigma_{\mathrm{el}} = \frac{G_B^2 }{\pi}A^2 \Enu^2 \simeq 1.7\times
10^{-38}\,\cm^2 \times A^2 \left( \frac{\N}{100} \right)^2 \left(
  \frac{\Enu}{10\,\MeV} \right)^2 .
\end{align}
The coefficient in front of the second relation serves as a figure of
merit when compared to the DM-nucleon cross section $\sigma_n$.  Given
that direct detection experiments have put upper limits on $\sigma_n $
as low as $10^{-44}\,\cm^2$~\cite{Aprile:2011hi} the coefficient
in~(\ref{eq:sigTotal}) is sizable.  However, a much more stringent cut
off in \Er\ makes it increasingly difficult for an essentially
massless $\nub$ to scatter off heavier targets.

The recoil spectrum arising from the solar flux of \nub\ will have to
include an average over the neutrino energy spectrum $df_i/d\Enu$ of
neutrino source $i$, weighted by the $\nub$ appearance probability and
an overall flux modulation $[L_0/L(t)]^2$ due to the earth's eccentric
orbit,
\begin{align}
\label{eq:rate}
  \frac{dR}{dE_R} = N_T \left[ \frac{L_0}{L(t)} \right]^2 \sum_i\Phi_i 
  \int^{\mathrm{E_{\mathrm{max},i}}}_{\mathrm{E_{\mathrm{min}}}} dE_{\nu}\, P_{eb}(t,E_{\nu})\frac{df_i}{dE_{\nu}}
  \frac{d\sigma_{\mathrm{el}}}{dE_R} . 
\end{align}
Note that $P_{eb}$ depends on $\Enu$ so that it has to be included
into the average; $\Phi_i$ is the integral flux given
in~(\ref{eq:fluxes}) and $df_i/dE$ is normalized to unity, \mbox{$\int
  dE\, df_i/dE=1$}, and taken
from~\cite{PhysRevC.54.411,PhysRevC.56.3391}. $N_T$ denotes the number
of target nuclei per unit detector mass, and in our computations we
take the fractional isotopic abundances of each element under
consideration into account.
The rate exhibits a non-trivial time-dependence. The maximum of the
overall flux is attained at the perihelion in early January. However,
the integral in Eq.~(\ref{eq:rate}) constitutes an additional source
of modulation which depends on the neutrino energy.  It will have its
strongest effect on the differential rate in the "just-so" regime of
Eq.~(\ref{eq:Losc}) where $L_{\mathrm{osc}}$ is on the order of the
sun earth distance; we exploit this fact in the following section.

\begin{figure}
\centering
\includegraphics[width=\textwidth]{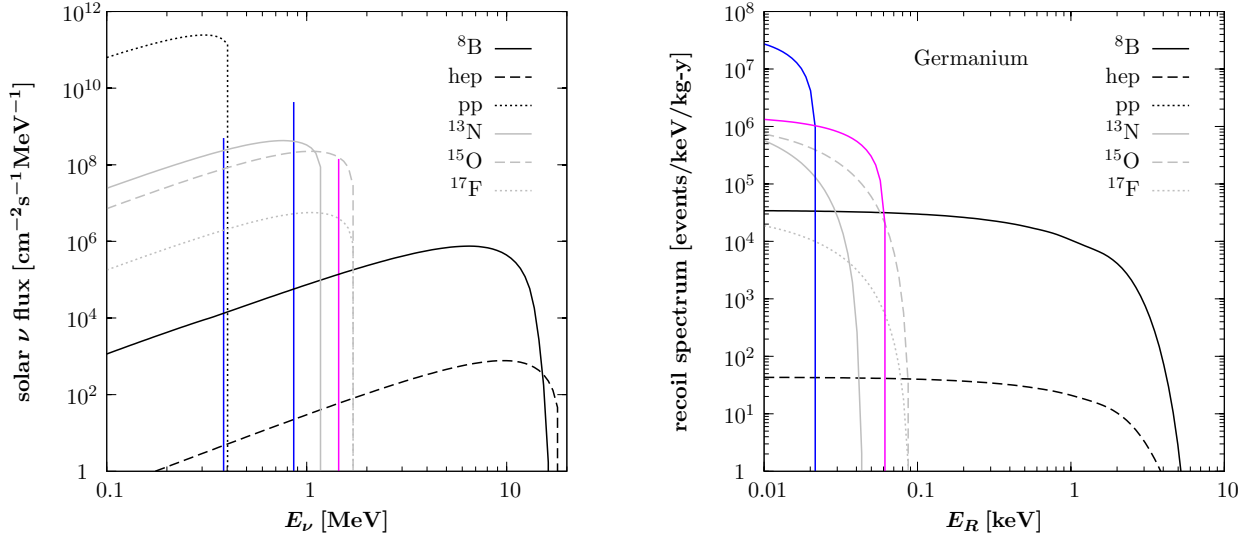} 
\caption{\small \textit{Left:} Solar neutrino fluxes as a function of
  energy as taken
  from~\cite{PhysRevC.54.411,PhysRevC.56.3391,PhysRevD.49.3923}. \textit{Right:}
  Associated recoil spectrum in a (perfect) germanium detector with a
  total exposure of 1~kg$\times$yr.  Only \bet\ and hep neutrinos
  reach out to values of $\Er$ where direct detection experiments
  become sensitive. }
\label{solnu}
\end{figure}

The left panel of Fig.~\ref{solnu} shows the solar neutrino spectra of
the various sources. In the right panel we compute the associated
recoil spectra using (\ref{eq:rate}) for a germanium detector without
threshold, with perfect energy resolution, and an exposure of
$1\,\kg\times$yr. For simplicity and since it does not affect the
argument we use $P_{eb}$ as given in (\ref{eq:Peb}) for all fluxes,
while strictly speaking Eq. (\ref{eq:Peb}) is only valid in the MSW
regime. The deviation from MSW only affects the softest recoils, and
can therefore be safely neglected in what follows.  As can be seen,
only \bet\ and hep neutrinos reach out to values of $\Er\gtrsim
\mathrm{few}\,\keV$ where direct detection experiments become
sensitive. Furthermore, \bet\ neutrinos constitute the dominant part
of the signal with hep giving a small correction only.

The spectrum in (\ref{eq:rate}) is a theoretical one. To make contact
with experiment we include effects from energy resolution, detector
threshold and quenching of nuclear recoil energy. Details will be
given when discussing the respective experiments. We start our
discussion by considering the experiments with putative positive
signal claims.

\subsection{DAMA}
\label{sec:dama}

The DAMA/NaI and its upgrade the DAMA/LIBRA
experiment~\cite{Bernabei:2008yh}, situated in the northern hemisphere
at the underground Gran Sasso National Laboratory (LNGS), were the
first to report on a potential direct DM detection.  The experiment
uses large radiopure NaI(Tl) crystals to measure scintillation light
resulting from nuclear recoils.  Given that there is no other
discriminating channel except requiring the candidate event to be a
``single-hit'', a relatively large overall count-rate of $\sim 1\,
\mathrm{cpd/kg/keVee}$ is observed. The presence of a positive signal
is inferred from the annual modulation of the residual count rate on
the order of $\sim 0.02\, \mathrm{cpd/kg/keVee}$ over low-energy bins
between 2 and 6\,\keVee\ once the average count rate per cycle is
subtracted~\cite{Bernabei:2008yi,Bernabei:2010mq}; for recent
discussions on potential modulating backgrounds
see~\cite{Ralston:2010bd,Nygren:2011xu,Blum:2011jf,Chang:2011eb,Bernabei:2012wp}.

\begin{figure}
\centering
\includegraphics[width=0.6\textwidth]{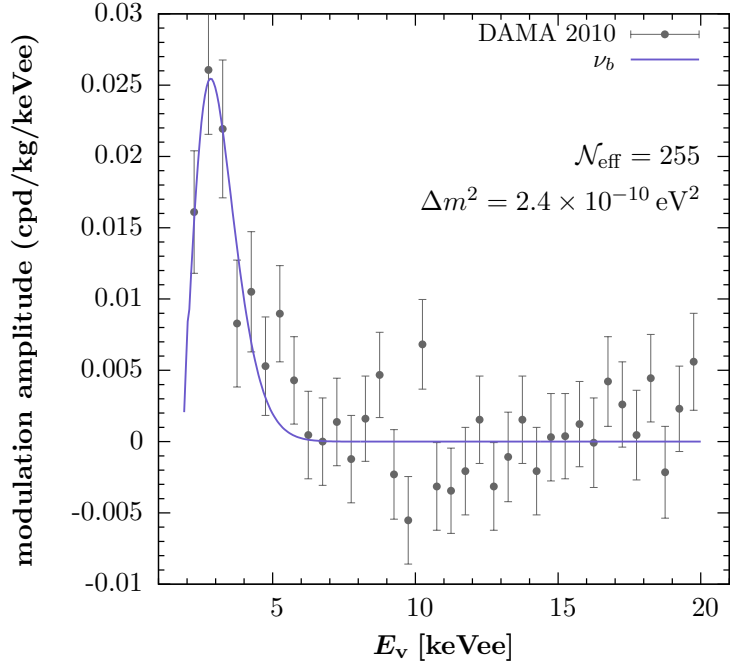}
\caption{\small The data points show the DAMA modulation amplitude as
  reported in~\cite{Bernabei:2010mq} in units of counts per day (cpd)
  per kg detector material and recoil energy. The solid line is the
  best fit from \nub\ to the data. }
\label{dama}
\end{figure}

\begin{figure}
\centering
\includegraphics[width=0.9\textwidth]{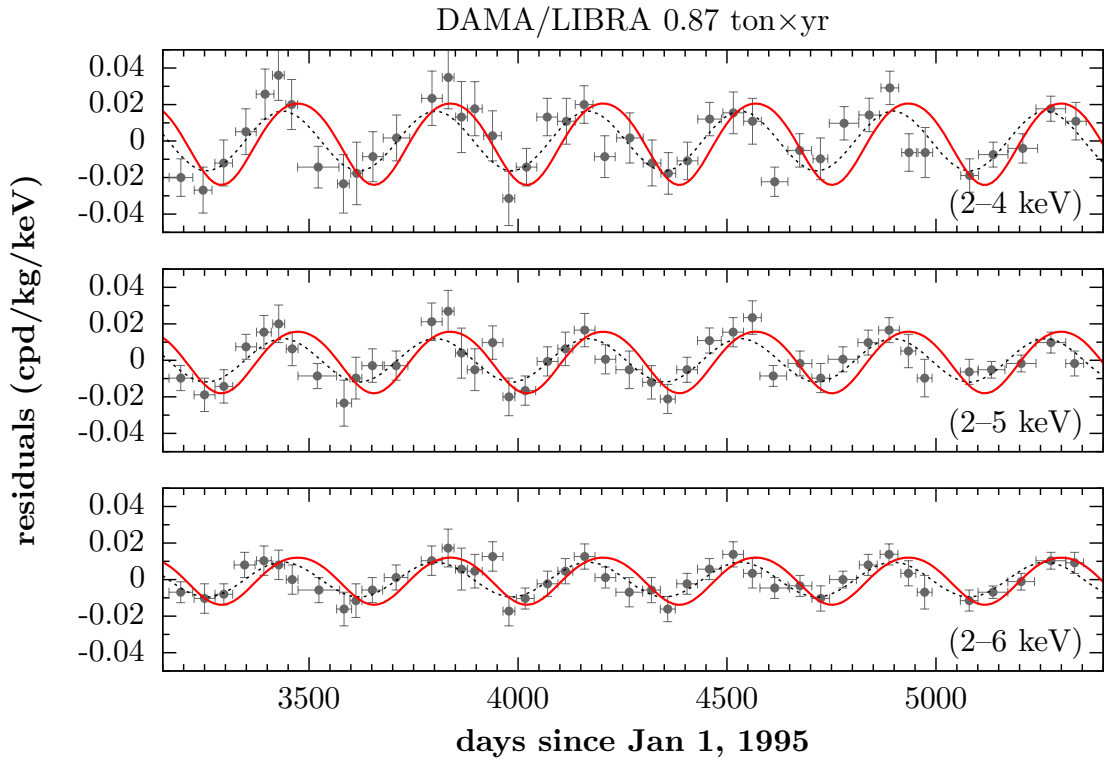} 
\caption{\small The data points show the DAMA/LIBRA reported residual event
  rate for various energy bins as a function of time. The red line is
  the residual event rate associated with the fit to the modulation
  amplitude in Fig.~\ref{dama}. The dotted line is a fit of the
  sinusoidal function $A\times \cos[\omega (t-t_0)]$ with $\omega$
  corresponding to a period of one year and a phase $t_0$ as expected
  from DM. As can be seen the \nub\ signal is approximately out of
  phase by one month. For a quantitative statement see main text.}
\label{residuals}
\end{figure}

The modulation of the event rate has been observed over the course of
more than a dozen annual cycles, collecting a cumulative exposure of
1.17\,ton$\times$yr~\cite{Bernabei:2008yi,Bernabei:2010mq}. The
null-hypothesis, \textit{i.e.} a rate constant in time, has been
excluded at the $8.9\sigma$ level. The residuals of the DAMA/LIBRA
runs in consecutive bins between (2-4)\,\keV, (2-5)\,\keV\ and
(2-6)\,\keV\ are shown by the data points in Fig.~\ref{residuals}. The
DAMA signal is usually decomposed as
\begin{align}
  S = S_0 + S_m \cos{\left[ \omega (t-t_0) \right]}
\end{align}
where $S_0 \sim 1\,\cpd/\kg/\keVee$ is the baseline rate of single hit
events and $S_m$ is the modulation amplitude,
\begin{align}
\label{eq:mod-amp}
  S_m = \frac{1}{2} \left( \left.\frac{dR}{d\Ev}\right|_{t_0}
    - \left.\frac{dR}{d\Ev}\right|_{t_0+1/2\,\mathrm{yr}} \right) .
\end{align}
The measured phase $t_0$ is reported to be compatible with the one
expected from DM, $t_0 = 152.5\,\days$ (June 2nd) with a period of one
year, \textit{i.e.} $\omega = 2\pi/(1\,\mathrm{yr})$. The reported
modulation amplitude is shown by the data points in Fig.~\ref{dama}. 

To see if \nub\ provides a viable explanation of the DAMA data one can
either fit the modulation spectrum (\ref{eq:mod-amp}) or directly the
time series of the residual rate. 
Considering the solar $\nub$ origin for DAMA, Eq.~(\ref{eq:rate}) may
not necessarily lead to a truly sinusoidal form of the signal as a
function of time. In addition, $t_0$ is not expected to be identical
with the DM value of 152.5\,days.  At first sight, a direct fit of the
time series seems therefore favorable.  However, the reported
residuals are binned in energy so that they only provide
coarse-grained information on the recoil energy distribution. This, in
contrast, calls for a fit of the modulation amplitude instead. We have
implemented both approaches and discuss their results below. In
addition, one can also attempt a joint fit of both data sets. This
approach is complicated by the fact that the data sets are not
independent.

We start by fitting  the modulation amplitude. Observable scatterings
of \nub\ occur on sodium only and no appreciable rate is expected for
$\Ev\gtrsim 7\,\keVee$. The latter expectation is in accordance with
what is seen in the data. Therefore, we only fit the first ten data
points with $\Ev\lesssim 7\,\keVee$ in order not to bias the
goodness-of-the-fit estimate. With the help of the usual $\chi^2$ function we
obtain the following best fit values,
\begin{align}
\label{eq:DAMA-bestFit}
\text{DAMA }S_m:\quad \dmb = 2.43\times 10^{-10}\,\eV^2,\quad
\Neff = 255,\quad \chi^2_{\mathrm{min}} /n_d = 9.5 / 8 .
\end{align}
The minimum in $\chi^2$ is associated with a $p$-value of $p=0.3$; $n_d$
denotes the number of degrees of freedom. The result of this fit is
shown by the solid line in Fig.~\ref{dama}. Confidence regions in
\dmb\ and \Neff\ are constructed by demanding,
\begin{align}
  \chi^2 ( \dmb, \Neff ) \leq \chi^2_{\mathrm{min}} + \Delta \chi^2 ,
\end{align}
where $ \chi^2_{\mathrm{min}} $ is given in
(\ref{eq:DAMA-bestFit}). We choose $ \Delta \chi^2 = 9.21$ which
corresponds to generous $99\%$~C.L.~regions. The choice results in the
two disjoint gray shaded regions shown in Fig.~\ref{fit}.

The above result looks promising. However, in contrast to the DM case
one has to check how well the time dependence of the rate is met. The
resulting signals in the various energy bins from the best fit
values~(\ref{eq:DAMA-bestFit}) are shown by the solid (red) lines in
Fig.~\ref{residuals}. For completeness we also show by the dotted
lines fits of the data by the sinusoidal function
$A_i\cos{[\omega(t-t_0)]}$ with period 1\,yr and
$t_0=\mathrm{June}\,2^{\mathrm{nd}}$ as expected when the origin were
due to DM scatterings.
As can be seen by eye, the \nub\ signals seems to lag behind by
approximately one month. Thus, the best fit corresponds to a phase
inversion with a maximum rate at the aphelion with $t_0\sim \text{July
  5th}$. 
For example, using the $(2-4)\,\keVee$ residuals, the best fit values
of the modulation spectrum (\ref{eq:DAMA-bestFit}), one finds
$\chi^2/n_d = 101/41$ with $p= 5\times 10^{-7}$ for the
residuals. This points towards a very poor description of the full
data.

We can try to improve on the above situation by directly fitting the
residual rate.  This is an important check, since the time dependence
of Eq.~(\ref{eq:rate}) is non-trivial. Can we find a corner in the
considered parameter space in which one can alleviate the tension in
the phase of DAMA and the \nub\ signal?
From the $(2-4)\,\keVee$ data we obtain as best fit,
\begin{align}
\text{DAMA residuals:}\quad \dmb = 2.45\times 10^{-10}\,\eV^2,\quad
\Neff = 183,\quad \chi^2_{\mathrm{min}} /n_d = 73.2 / 41 .
\end{align}
Though the fit fares slightly better on the residuals with $p\simeq
10^{-3}$, this does not ameliorate the situation sufficiently.
Moreover, the smaller value of \Neff\ now somewhat under-predicts the
modulation amplitude. Finally, even when we perform a joint fit
(neglecting potential covariances of the data sets) we do not find any
substantial improvement. We conclude, that even though the DAMA
modulation amplitude is fit rather nicely, the time series of events
speaks against the \nub\ interpretation.

As a final remark, we comment on the sodium quenching factor. For our
analysis above we used $Q=0.3$ in the conversion to electron
equivalent recoil energy, $\Ev(\keVee) = Q \Er (\keV)$. New
measurements~\cite{UCLAJuan} seem to indicate 1) lower values $Q\sim
0.15 $ and 2) a stronger energy dependence as previously thought. This
has important implications for light DM as well as for the \nub\
hypothesis. We find that $Q=0.15$ moves the DAMA regions in
Fig.~\ref{fit} towards larger values of \Neff\ which are already
excluded by all the other considered null searches. The situation then
becomes more similar to the one already witnessed for DM.

\subsection{CoGeNT}
\label{sec:cogent}

The CoGeNT experiment is a low-threshold nuclear recoil germanium
detector situated in the Soudan Underground Laboratory. The latest
data release is based on 442 live days taken with 0.33~kg
target~\cite{Aalseth:2010vx,Aalseth:2011wp}. An unexplained
exponential rise of the signal at lowest recoil energies
0.5--1\,\keVee\ is observed. The origin of it is unknown and has lead
to the speculation that DM with a mass in the $\sim 8-10\,\GeV$ range
may be the cause of it. For spin-independent DM-nucleus scattering,
the excess requires a cross section $\sigma_{SI}\sim
10^{-40}\,\cm^2$. Such values are challenged by the null-result of
XENON100 and by the low-threshold analysis of CDMS-II. However, more
recently the collaboration has identified a source of
surface-background events which may lead to a revision of the signal
strength in the low-recoil bin 0.5--1\,\keVee. In the following we
will investigate the possibility that the observed excess may be due
to scattering of \nub\ in the detector. We will also account for the
possibility that the collaboration's results could be revised in the
near future~\cite{UCLAJuan}.

In addition to the signal-rise below 1\,\keVee\, the data also appears
to be annually modulated in the 0.5--3.2\,\keVee\ bracket. The
observed event rate peaks in mid-to-late April (2010) with a
modulation amplitude of $\sim 16\%$, most pronounced between
1.4--3.2\,\keVee~\cite{Aalseth:2011wp}. The latter behavior is neither
expected from DM scatterings nor could it be explained by \nub\
scatterings since the recoil spectrum arising from \bet\ neutrinos is
cut off for $\Ev \gtrsim 1.4\,\keVee$. We also note that the
modulation of nuclear recoil events in Ge in that energy regime has
recently been challenged in a dedicated analysis by
CDMS~\cite{UCLAcdms}.  We will therefore not further address the
potential modulation of the CoGeNT signal and await further data.

\begin{figure}
\centering
\includegraphics[width=0.6\textwidth]{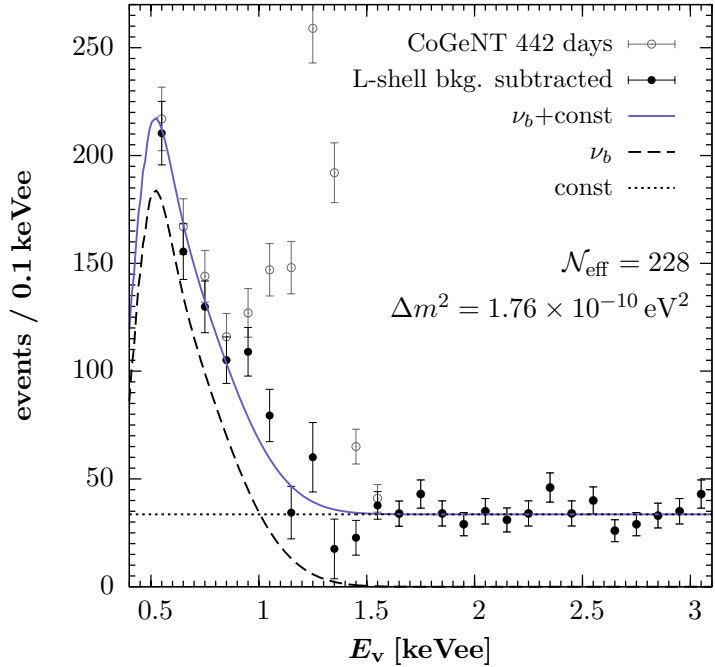} 
\caption{\small Recoil spectrum from the 442 live-day run of the
  CoGeNT experiment. The black (gray) data points show the signal
  after (before) subtraction of the cosmogenic radioactive
  background. The solid line is a fit to the black data points. It
  decomposes into the contribution from \nub\ (dashed line) and the
  contribution of a constant background (dotted line.)}
\label{cogent}
\end{figure}

Cosmogenically induced radioactive background has to be subtracted
from the CoGeNT data before fitting the exponential excess.
The radioimpurities in the crystal have been identified by the
collaboration, with the most prominent ones given by the electron
capture decays of $^{68}$Ge and $^{65}$Zn centered at 1.3\,\keVee\ and
1.1\,\keVee, respectively.
From a fit of observed K-shell electron capture peaks seen in the high
energy data and from the expected ratio of L-to K-shell decays one can
subtract the low-energy L-shell background in the 0.5--3.2\,\keVee\
window. We follow~\cite{CogentWriteUp} in the subtraction and collect
the time-stamped events in 0.1\,\keVee\ bins.  The result of this
procedure can be seen in Fig.~\ref{cogent} as the difference between
gray (with peaks) and black (peaks subtracted) data points.

Nuclear recoil energies on germanium have to be converted into the
measured ionization signal. We employ a Lindhard-type, energy
dependent quenching factor, $\Ev(\keVee) = Q \times
\Er(\keV)^{1.1204}$ with $Q = 0.19935$~\cite{CogentWriteUp} and
account for a finite detector resolution by convolving the recoil
signal with a Gaussian of width $\sigma^2 = (69.4\,\eV)^2 +
0.858\,\eV\times\Ev(\eV)$~\cite{Aalseth:2008rx}. Finally, the
efficiency of the detector is provided in Fig.~1
of~\cite{Aalseth:2011wp}.

When fitting the CoGeNT excess at low energies we follow two
approaches: In the first case we seek an explanation of the excess
exclusively in terms of \nub\ scatterings on Ge, allowing only for a
constant background contribution. In the second case we relax the
assumption on the background and allow, in addition, for an
exponential background component, $A\times \exp(-B \Ev)$, with
coefficients $A$ and $B$ to be determined in the fit. Clearly, in the
latter approach a \nub-induced contribution may not even be necessary
as the excess resembles an exponential shape. This is therefore the
most conservative way to treat the data in terms of new physics and it
will show us the ``compatibility'' region in the (\Neff,$\Delta m_b$)
parameter space. As mentioned above, the CoGeNT excess is likely to be
revised by the collaboration. In the second approach, the additional
exponential background is here to mimic that possibility without
quantifying its (yet unknown) concrete strength.

We first take the reported CoGeNT excess at face value and fit it by
\nub\ scatterings on Ge together with a constant background rate. The
best-fit values are
\begin{align}
\label{eq:cogent-bestfit}
\text{CoGeNT:}\quad \dmb = 1.76\times 10^{-10}\,\eV^2,\quad
\Neff = 228,\quad \chi^2 /n_d = 33.6/24 ,
\end{align}
which corresponds to $p\simeq 0.1$. The background rate is $c_0 =
3.36\,\cpd/\kg/\keVee$. A finer-grained bin size improves the
goodness-of-fit to $\chi^2 /n_d = 47.6/47$ with $p=0.45$. We consider
this a very good description of the data. One should keep in mind that
the subtraction of the cosmogenic background has uncertainties itself
which are not accounted for in the errorbar. Figure~\ref{cogent}
shows the spectrum obtained from (\ref{eq:cogent-bestfit}).  The
dashed line shows the signal from \nub\ only and the solid line
includes the constant background. Below $\Ev\lesssim 0.5\,\keVee$ the
detector efficiency decreases rapidly, which explains the turn-off of
the scattering signal.

Figure~\ref{fit} shows the inferred 99\%~C.L.~regions in the
$(\dmb,\Neff)$ parameter space as labeled. We use $\Delta \chi^2=
9.21$, \textit{i.e.}~we treat the constant background rate as a
nuisance parameter. This is equivalent of using a profile likelihood
to infer the confidence regions. Two isolated islands are visible as
labeled. The thin gray solid line labeled ``CoGeNT hull'' which
touches the CoGeNT regions from above is obtained by allowing an
additional exponential background in the fit (see discussion above.)
Without further knowledge of the strength of this background, the full
region below the line then becomes viable. Whenever the \nub\ signal
becomes too weak, the background takes over in producing a viable
fit. The general expectation is that once the collaboration revises
their statements about the strength of the exponential rise, the
CoGeNT favored regions will move vertically downwards, but at this
point it is impossible to speculate by how much. In conclusion, we
find that \nub\ can provide an excellent explanation to the CoGeNT
data.

\subsection{CRESST-II}
\label{sec:cresst-ii}

The CRESST-II experiment~\cite{Angloher:2004tr},
situated at LNGS, has recently presented their results from their DM
``run32'' with a total of 730~kg$\times$days effective exposure
between 2009-2011~\cite{Angloher:2011uu}. The analysis has been
carried out using data collected by eight CaWO$_4$ crystals which
measure heat and scintillation light resulting from nuclear
recoils. The calorimetric phonon channel allows for a precise
determination of the recoil energy with a resolution better than
1\,keV. Nuclear recoils are again quenched in scintillation
light. This is a virtue as it allows for a discrimination against
$e^{-}$ and $\gamma$ induced events. Moreover, the quenching factors
of Ca, W, and O differ. To a limited degree, recoils against the
respective elements can therefore be distinguished.

\begin{figure}
\centering
\includegraphics[width=0.6\textwidth]{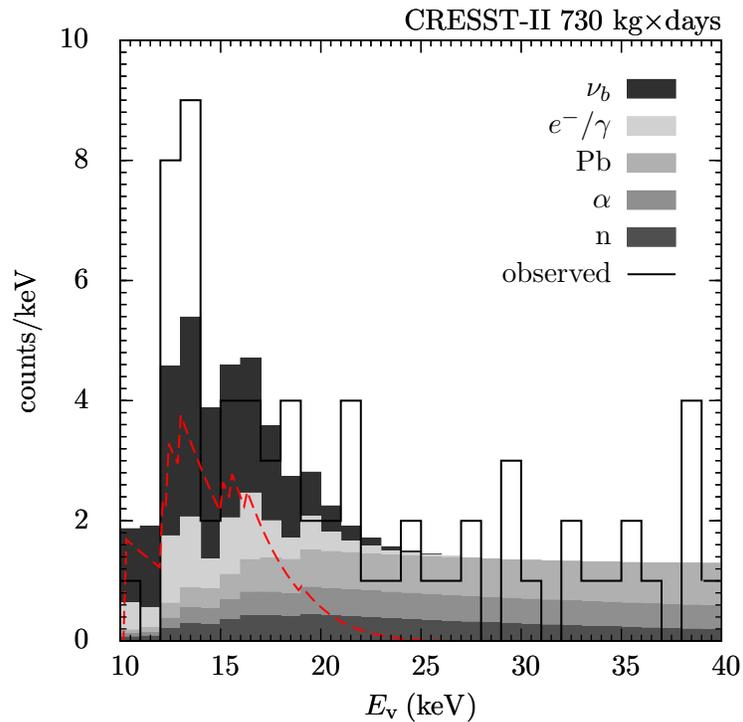} 
\caption{\small CRESST-II recoil spectrum. The solid line is the
  histogram of reported events in the $730\,\kg\times\days$ run
  summing to a total of 67 events. The gray shaded (stacked)
  histograms show the best fit contribution from $\nub$ (darkest
  shading) and the modeled backgrounds as labeled and explained in the
  main text. The spiky dashed (red) solid line is the unbinned \nub\
  signal. }
\label{cresst}
\end{figure}

The analysis~\cite{Angloher:2011uu} finds an intriguing accumulation
of a total of 67~events in their overall acceptance region between
10--40\,keV, shown by the solid line in Fig.~\ref{cresst}. The
low-energy threshold of each detector-module is determined by the
overlap between $e/\gamma$- and nuclear recoil band. Allowing for a
leakage of one background $e/\gamma$-event per module distributes the
individual detector thresholds between 10.2--19\,\keV. Whereas
$e/\gamma$-events are a well controllable background, the experiment
suffers from a number of less well-determined sources of spurious
events. To assess how well the observed events can be explained in
terms of new physics makes the modeling of such
background---unfortunately---unavoidable.

In the following we briefly mention each of the known background
sources and outline our treatment of them:
\begin{enumerate}
\item As alluded before, the thresholds of the detector modules are
  chosen such that a leakage of a total number of 8 $e/\gamma$-induced
  events into the nuclear bands are expected. We find the energy
  distribution of these events by digitizing and binning the
  corresponding line from Fig.~11 of~\cite{Angloher:2011uu}.
\item Degraded $\alpha$-particles from radioactive contamination in
  the clamping system holding the crystals can be misidentified as
  nuclear recoils once their energy falls below 40\,\keV.  A sideband
  analysis above that energy indicates that the distribution in recoil
  energy is flat. This allows to estimate the number of
  $\alpha$-events in the acceptance region for each detector module
  and which is provided in Tab.2 of~\cite{Angloher:2011uu}. We follow
  this prescription which yields a total of 9.2~events.
\item Related to the previous source, $^{210}$Pb $\alpha$-decays from
  radioactive lead on the clamps holding the crystal and with the
  $\alpha$-particle being absorbed by non-scintillating material
  constitutes another source of background. The peak at 103\,\keV\
  full recoil energy in $^{206}$Po is clearly visible and a fit of it
  allows to infer the overall exponential tail distribution in the
  acceptance region below 40\,\keV. The low-energy tail is due to
  $^{206}$Po that is slowed down in the clamps before interacting in
  the crystal.  We estimate the radioactive lead contamination of each
  detector module from the number of observed events in the reference
  region above 40\,\keV. This yields a nominal background of about
  17~events.
\item Finally, low-energy neutron-nucleus scatterings in the crystals
  is a well known source of background. These neutrons can be produced
  by in-situ radioactive sources as well as by cosmogenic muons in-
  and out-side the detector housing. Since neutrons tend to scatter
  more than once, some information on the overall flux can be obtained
  from the amount of coincident events in different detectors. Such
  multiple scatters also tend to wash out the initial spectral
  information. The authors of~\cite{Angloher:2011uu} parameterize the
  neutron spectrum by a simple exponential $dN_n/dE = A \times
  \exp(-E/E_{\mathrm{dec}})$ where $E_{\mathrm{dec}}=(23.54\pm
  0.92)\,\keV$ has been obtained from a neutron calibration run with
  an AmBe source. The best we can do is assuming a uniform neutron
  flux in all detector-modules. With $A=1$ one gets a total of about 9
  events. When fitting \nub\ to the data we leave $A$ as a free
  parameter. We observe that $A$ is never too large once the goodness
  of the fit becomes acceptable. The reason is that the \nub-induced
  spectrum experiences are relatively sharp cutoff for energies in
  excess of $\sim 25\,\keV$. Thus, the high-energy part of the
  acceptance region has to be entirely explained by background for
  which $A =\mathcal{O}(1)$ provides the best fit.  \nub\ scatters
  mainly on oxygen and calcium, so that there is no need to further
  dissect the neutron background as the latter also scatters to 90\%
  on O~\cite{Angloher:2011uu}.
\end{enumerate}

In~\cite{Angloher:2011uu} the various nuclear recoil bands have not
been resolved.  Therefore, we compute rate predictions for CRESST
summing up all events in Ca, O, and W.
The fractional exposures of the individual detector modules lie within
$\sim 20\%$ of a uniform one with value $1/8$. We use the accurate
values as provided by~\cite{Jens}. We account for a finite Gaussian
energy resolution in the phonon channel with $\sigma =1\,\keV$. Since
the number of observed events $n_i$ in each of the bins is small, we
fit the data by minimizing the Poisson log-likelihood ratio
\begin{align}
\label{eq:poisson-L}
  \chi^2_P = 2 \sum_{i} \left[ y_i - n_i + n_i \ln{\left(
        \frac{n_i}{y_i} \right)} \right] ,
\end{align} 
where the sum runs over all bins and $y_i$ is the sum of background
and signal contributions; the last term is absent when $n_i=0$.
Confidence regions are directly constructed from~(\ref{eq:poisson-L}).

Figure~\ref{cresst} shows the recoil spectrum induced by $\nub$ as the
(magenta) continuous and falling line. Summing all background
contributions to the \nub\ signal the best fit parameters read,
\begin{align}
\label{eq:cresst-bestfit}
\text{CRESST-II:}\quad \dmb = 3\times 10^{-10}\,\eV^2,\quad
\Neff = 49,\quad \chi^2 /n_d = 27.7/27 ,
\end{align}
with a $p$-value $p=0.48$ under the approximation that $\chi^2_P$ in
(\ref{eq:poisson-L}) follows a $\chi^2$ distribution with $n_d =
27$. The amplitude of the neutron background is found to be
$A=1.23$. Fixing instead $A=1$ yields the same parameters
(\ref{eq:cresst-bestfit}) with negligible degradation in $\chi^2$.
Discontinuous jumps in the count rate when going from lower to higher
recoil energies are due to the onsets of the various detector modules
with increasing energy thresholds. Also shown as a stacked histogram
are the modeled sources of background as labeled. 

\begin{figure}
\includegraphics[width=\textwidth]{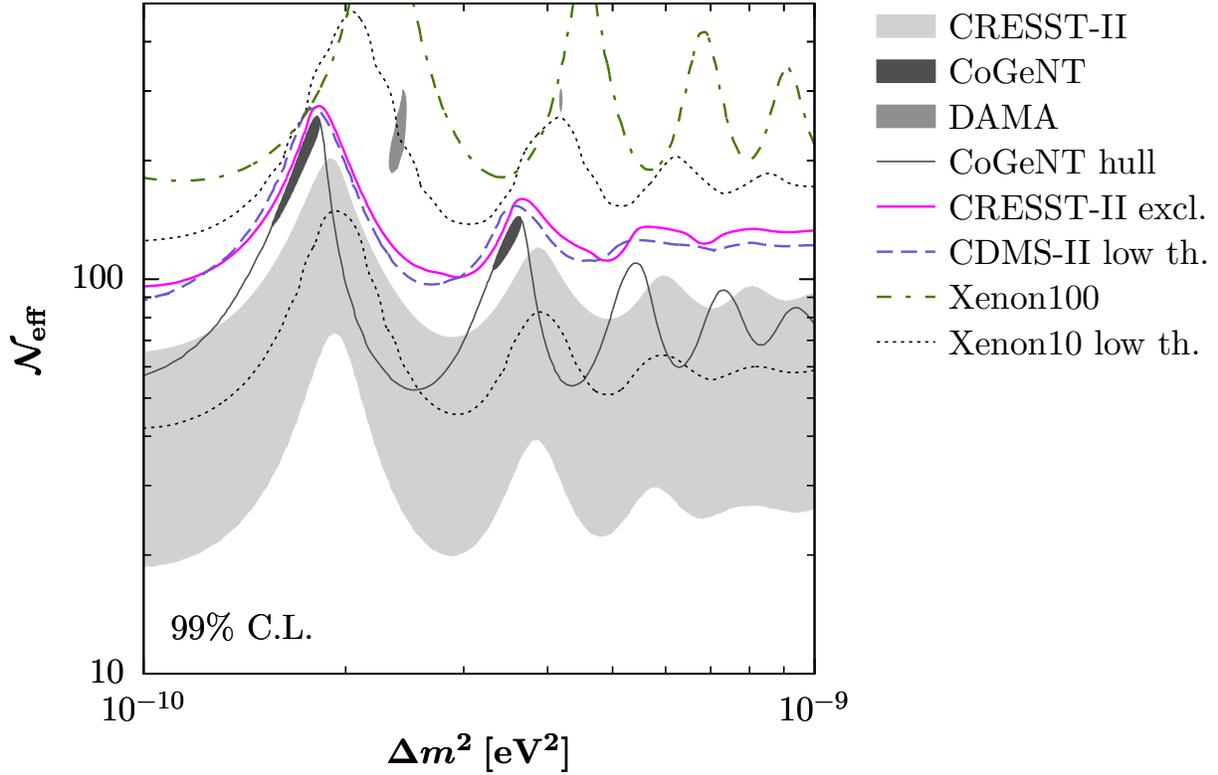} 
\caption{\small Summary plot of direct detection favored regions and
  constraints in the parameters \dmb\ and \Neff\ at 99\%
  confidence. \textit{Favored regions:} the broad light shaded gray
  band shows the CRESST-II region. The two darkest islands are the
  regions in which the CoGeNT excess is explained. In presence of an
  exponential background contamination (\textit{e.g.} due to ``surface
  events''), the region below the thin gray line labeled as ``CoGeNT
  hull'' becomes in principle viable (see main text for details.) The
  two medium gray shaded islands indicate the regions in which the
  DAMA modulation amplitude is fitted; these regions as well as any
  other parameter choices however exhibit a tension in timing when
  compared to the DAMA residuals. \textit{Constraints:} \Neff\ values
  above the respective lines are excluded (or seriously challenged.)
  The top constraint is the one from Xenon100, the two degenerate ones
  below are obtained with the CRESST-II data and CDMS-II low-threshold
  data. The two dotted lines at the bottom show the constraints
  arising from the Xenon10 low-threshold analysis with two different
  assumptions on the ionization yield $Q_{y}$ (see main text for
  details.)}
\label{fit}
\end{figure}

Figure~\ref{fit} shows again the 99\% confidence regions in the
(\dmb,\Neff) parameter space. As can be seen, the favored region
stretches across the plane and the trend for larger values of $\dmb$
not shown in the plot can be easily be extrapolated. The CRESST region
is compatible with the one inferred from the DAMA modulation
amplitude. Once the CoGeNT excess is revised (see previous section),
it is very likely that the resulting best fit region will overlap with
CRESST too. CRESST spans a rather wide region in parameter space.  The
light yield distribution of the candidate events as a function of
$\Ev$ is not published and has thus not been accounted for. With
eventual better knowledge of this quantity, the region is expected to
shrink in a joint fit. In the \nub\ scenario, most of the scatters are
on oxygen which in turn yields most scintillation light among the
CaWO$_4$ constituents.

It is also likely that the new CRESST data constrains larger values of
$\Neff$. This is especially true given that the detector model with
lowest threshold only observed one count between 10--12\,\keV. In
order to put a constraint we use what has been termed ``binned
Poisson'' technique in~\cite{Savage:2008er}: for one bin, an average
number of events $\nu = \nu_{\mathrm{s}} + \nu_{\mathrm{bg}}$
consisting of $ \nu_{\mathrm{s}} $ signal and $ \nu_{\mathrm{bg}} $
background events is excluded at a level $1-\alpha_{\mathrm{bin}}$ if
the probability to see as few as $n_{\mathrm{obs}}$ observed events is
$\alpha_{\mathrm{bin}}$. Since $n_{\mathrm{obs}}$ is a Poisson
variable, $\alpha_{\mathrm{bin}}$ is given the lower tail of the
Poisson distribution, $ \alpha_{\mathrm{bin}} =
\sum_{n=0}^{n_{\mathrm{obs}}} \nu^n \exp(-\nu)/n! $. When dealing with
more than one bin the level of exclusion $1-\alpha$ is given by
\begin{align}
  \label{eq:confExcl}
1-\alpha  = (1-\alpha_{\mathrm{bin}})^{N_{\mathrm{bin}}} ,
\end{align}
where $N_{\mathrm{bin}}$ is the number of bins; $\alpha$ is the
probability to see as few events as observed in at least one of the
bins. For placing a constraint from CRESST we only use the bins from
10--25\,\keV\ as those are the ones for which $\nub$ can give a
contribution. More conservative constraints are obtained when assuming
that there is no background.

\subsection{Null-searches}
\label{sec:null-searches}

\subsubsection{CDMS-II low threshold analysis}
\label{sec:cdms-ii-low}

The CDMS-II collaboration has published a low threshold analysis from
the Soudan site using eight Ge detectors with a raw exposure of
241\,\kg\,\days~\cite{Ahmed:2010wy}.  At the expense of discriminating
power of $e^{-}/\gamma$ against nuclear recoils, a threshold of
$2\,\keV$ was reached. This is an interesting analysis because it uses
the same target material as CoGeNT. Indeed, for the DM case the
results indicate a serious conflict between the two experiments.
Therefore, it is important to check if the CoGeNT explanation is also
challenged in the baryonic neutrino scenario.

An exponential-like signal rise towards threshold with a maximum event
rate of $\sim 1\,\cpd/\kg/\keV$ has been observed; see Fig.~1
in~\cite{Ahmed:2010wy}.  ``Zero-charge'' events from electron recoils
near the edge of the detector are expected to yield a major
contribution to the observed count rate. Given that the estimation of
this background involves extrapolation and hard-to-control systematic
errors, we obtain the most conservative limits on \Neff\ by not
subtracting this background. This follows the approach taken by the
collaboration itself.%
\footnote{It has been speculated~\cite{Collar:2011kf} that the CoGeNT
  and CDMS-II recoil spectra are indeed similar after correcting for a
  potential energy miscalibration. Given that the status of the CoGeNT
  recoil spectrum is uncertain itself, we do not follow up on that
  discussion in this work. }
Only the first three data bins covering $2\,\keV \leq \Er \leq
3.5\,\keV$ are sensitive to \nub\ scattering. On those bins we perform
a ``binned Poisson'' exclusion, similar to the one explained in
Sec.~\ref{sec:cresst-ii}. We correct for efficiency according to
Fig.~1 in~\cite{Ahmed:2010wy} and use a Gaussian detector resolution
of $0.2\,\keV$.  In Fig.~\ref{fit} we show the resulting constraint at
99\%\,C.L. Remarkably, CDMS-II does not challenge the CoGeNT-favored
regions.

\subsubsection{SIMPLE}
\label{sec:simple}

An interesting direct detection experiment in the current context is
SIMPLE~\cite{Felizardo:2011uw}, operated in the Low Noise Underground Laboratory in southern
France. It uses a dispersion of superheated liquid droplets made of
C$_{12}$ClF$_{5}$ with an total active mass of 0.2\,\kg. Only nuclear
scattering induces phase transition which results in bubble
nucleation. The fact that the experiment contains mainly light
elements makes it susceptible to \nub-scattering. Notably, fluorine
with atomic number $A=19$ has a target mass fraction of $\sim 60\%$.

We use the results from Phase I of Stage II with 14.1\,\kg\,\days\
exposure~\cite{Felizardo:2011uw}. A total of 8 events were observed
with an expected neutron background of 12. We include this background
in the derivation of an upper limit of \Neff\ as a function of \dmb.
We model bubble nucleation and heat transfer efficiency
following~\cite{Felizardo:2011uw} for which we use an energy threshold
of $8\,\keV$. For better overview, we did not include the constraint
in Fig.~\ref{fit}. It is superseded by the ones from CDMS-II and
CRESST-II but more constraining than Xenon100 as will be discussed
next.

\subsubsection{Xenon experiments}
\label{sec:xenon}

In this work we consider the results from the Xenon10 and its upgrade,
the Xenon100 experiment at LNGS~\cite{Aprile:2010bt,Aprile:2011hi}.
Currently, the most stringent constraint for spin-inelastic scattering
for DM masses in the $50\,\GeV$ ballpark is the one from the
Xenon100~\cite{Aprile:2011hi}. The experiment also provides strong
limits in the light-DM mass region $\mathcal{O}(10\,\GeV)$. In the
latter context, the low-threshold analysis~\cite{Angle:2011th} from
Xenon10 is of particular interest.

The experiments use a prompt scintillation light signal (S1) and a
delayed one from ionization (S2) to detect the nature of the recoil
event. Both signals are measured in units of photo-electrons (PEs).
Nuclear recoil energies are obtained from the respective signals via
\begin{align}
\label{eq:LeffEr}
 S1:\: \Er = \frac{1}{\mathcal{L}_{\mathrm{eff}}} \frac{\mathrm{S}1}{L_y}
  \frac{S_e}{S_n}, \qquad S2:\:  \Er = \frac{\mathrm{S}2/\zeta}{Q_y} .
\end{align}
Only a small fraction of the deposited recoil energy is emitted in
form of scintillation light. The crucial quantity is the scintillation
efficiency $\mathcal{L}_{\mathrm{eff}}$ which it determines the
discrimination threshold of the experiment. Therefore, for Xenon100 we
will mostly be interested in S1 since a low-threshold analysis has not
yet been published. Conversely, for Xenon10 we focus on S2. For S1 we
use the measurements of~\cite{Plante:2011hw} with a conservative
extrapolation of $\mathcal{L}_{\mathrm{eff}}$ to zero at $2\,\keV$
nuclear recoil.  $L_y = 2.2 $ (Xenon100) is the light yield in
PEs/keVee obtained from $\gamma$-calibration. $S_e$ and $S_n$ are
quenching factors for scintillation light due to electron and nuclear
recoils, respectively. They are experiment-specific and depend on the
applied drift voltage; $S_e = 0.58$ and $S_n=0.95$ for Xenon100.
$Q_y$ is the ionization yield per keV nuclear recoil and $\zeta$ is
the measured number of PEs produced per ionized electron, $\zeta =
20\, (24)\,$PEs for Xenon100 (Xenon10); we will comment further on
$Q_y$ below.

The S1 detector acceptance for Xenon100 is found from the lines
presented in Fig.~2 of~\cite{Aprile:2011hi} together with a low-energy
threshold of 4\,PE which corresponds to $8.4\,\keV$ nuclear recoil
energy. For S2, the trigger efficiency is effectively 100\% in both
experiments.  We take the Poissonian nature on the expected number of
scintillation photons/ionization electrons into account. For example,
for S2 one computes
\begin{align}
  \frac{dR}{dn_e} = \int_{E_{\mathrm{min}}} d\Er\, \frac{dR}{d\Er} \times
  \mathrm{Poiss}(n_e|\nu_e(\Er)) ,
\end{align}
where $n_e = \Er Q_y$ is the number of ionized electrons; a
PMT resolution of 0.5\,PE can be neglected when converting $n_e$ to
S2; $\mathrm{S}2 = n_e\zeta$. We use a hard cut off at $E_{\mathrm{min}} = 1.4\,\keV$
following~\cite{Angle:2011th}.

We first discuss the constraint from
Xenon100. Reference~\cite{Aprile:2011hi} presents the results from a
100\,day run with a fiducial detector mass of 48\,\kg. Three events
were observed in the acceptance region $8.4$--$44.6\,\keV$. We use
Yellin's maximum gap method~\cite{Yellin:2002xd} to set an upper
limit. The resulting constraint is again shown in
Fig.~\ref{fit}. Given that Xe is a rather heavy target and that the
scintillation threshold is $\mathcal{O}(10\,\keV)$, the constraint on
the parameter space is very week. Indeed, none of the favored regions
is challenged.

In contrast, a more stringent limit can be expected from the Xenon10
low-threshold analysis in~\cite{Angle:2011th}. Discarding the
scintillation signal allows one to lower the threshold to $\Er =
\mathcal{O}(1\,\keV)$.
After all cuts are applied and with a resulting effective exposure of
$\sim 6.15\,\kg\,\days$ a mere number of seven events in the region $
1.4\,\keV\lesssim \Er\lesssim 10\,\keV$ are observed. The calibration
of the nuclear energy recoil scale now solely depends on $Q_y$ [see
Eq.~(\ref{eq:LeffEr})]. The problem in the analysis is that one
requires $Q_y$ in a regime where no data is available and the
extrapolation of that quantity is not well supported by theoretical
expectations. Indeed, the adopted values in~\cite{Angle:2011th} by the
Xenon10 collaboration have been repeatedly
disputed~\cite{Collar:2010ht,Collar:2011wq}.
Therefore, we adopt two different extrapolations of $Q_y$. First we
use the solid line in Fig.~2 of~\cite{Angle:2011th} and second we
employ more conservative choice discussed in~\cite{Collar:2011wq} in
the estimation of the energy scale.

The resulting constraints are again presented in Fig.~\ref{fit}. In
contrast to all other null searches, Xenon10 can challenge the
entire region of interest with $\Neff\gtrsim 40$. We caution the
reader that it is also the constraint with the largest uncertainty. We
will further illustrate the sensitivity of the results on
$Q_{y}$ below when considering a prospective Xenon100 low-threshold
analysis.

\subsection{Future sensitivity in Direct Detection}
\label{sec:future-sens-direct}

\subsubsection{Xenon100 low threshold}
\label{sec:xenon100-low-thresh}

In the previous section we have seen that a low-threshold analysis in
Xenon10 can yield very stringent constraints. Therefore, it is
conceivable that the collected 100 live days of data with the Xenon100
detector may offer another sensitive test for this model. Here we
present projections for a Xenon100 ionization-only (S2) study.

\begin{figure}
\centering
\includegraphics[width=0.5\textwidth]{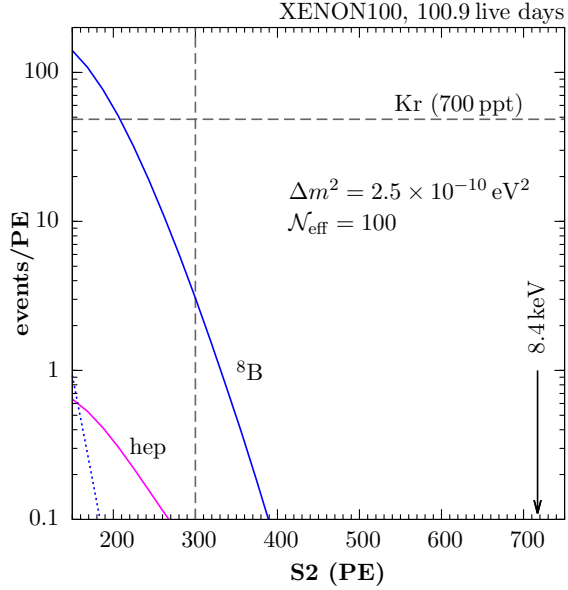} 
\caption{\small Projection for a Xenon100 low-threshold analysis for
  an exemplary parameter choice $\dmb = 2.5\times 10^{-10}\,\eV^2$ and
  $\Neff = 100$. The $x$-axis gives the ionization signal S2 in units
  of PEs. The horizontal and vertical dashed lines show the maximum
  rate from radioactive Kr decay and the S2 software threshold of the
  detector, respectively. The solid lines are the \nub-signals from
  \bet\ and hep neutrinos as labeled. The vertical arrow at 700~PEs
  indicates the current threshold of the S1 scintillation signal. The
  dotted line shows again \bet\ neutrinos for a calibration scale
  following \cite{Collar:2011wq} instead of \cite{Angle:2011th}. This
  highlights the severe sensitivity on the extrapolation of $Q_y$. }
\label{xe100} 
\end{figure}

Once the prompt scintillation signal S1 is discarded, the goal lies in
lowering the threshold in S2 as much as
possible. In~\cite{Aprile:2011hi} the Xenon100 quoted software
threshold is 300\,PE which corresponds to 20 ionized electrons. This
is a factor of five larger compared to the Xenon10 low-threshold
analysis. The gain in exposure by about an order of magnitude somewhat
compensates for this since the increase in detector mass does not
seriously affect the extraction efficiency of ionized
electrons. However, an air-leak during the run introduced unwanted Kr
contamination at the ($700\pm 100$)\,ppt level. The associated rate in
the electron recoil band reads,
\begin{align}
\label{eq:kr}
  R_{\mathrm{Kr}}< 22 \times 10^{-3}\: \cpd/ \kg/ \keVee , 
\end{align}
and is expected to be homogeneously distributed over recoil
energy. Clearly, a putative \nub\ signal must overcome this
background. We remark in passing that for future runs (\ref{eq:kr})
will diminish since the $^{85}$Kr concentration is being continuously
reduced by cryogenic distillation.

Figure~\ref{xe100} shows the sensitivity to \nub\ with the current
dataset of 100 live days for $\dmb = 2.5\times 10^{-10}\,\eV^2$ and
$\Neff = 100$. We present the recoil spectrum as a function of actual
detected S2 in units of PE. The solid lines show the \bet\ and hep
neutrino spectrum as labeled. The vertical dashed line indicates the
Xenon100 threshold and the horizontal one is the rate (\ref{eq:kr})
from Kr contamination.
The dotted line is again a \bet\ spectrum but this time with a
calibration scale following~\cite{Collar:2011wq}. As a consequence the
signal falls entirely below the threshold. 
Even a small change $\Delta \Er$ in the nuclear recoil calibration has
a large effect since $\Delta \mathrm{S}2 \sim \zeta \Delta\Er$.
This illustrates 1) that the Xenon10 constraints in the previous
section should be viewed with care and 2) that without further
experimental insight into $Q_y$ in (\ref{eq:LeffEr}) a conclusive
prediction for Xenon100 is not feasible.

\subsubsection{COUPP}
\label{sec:coupp}

\begin{figure}
\centering
\includegraphics[width=0.5\textwidth]{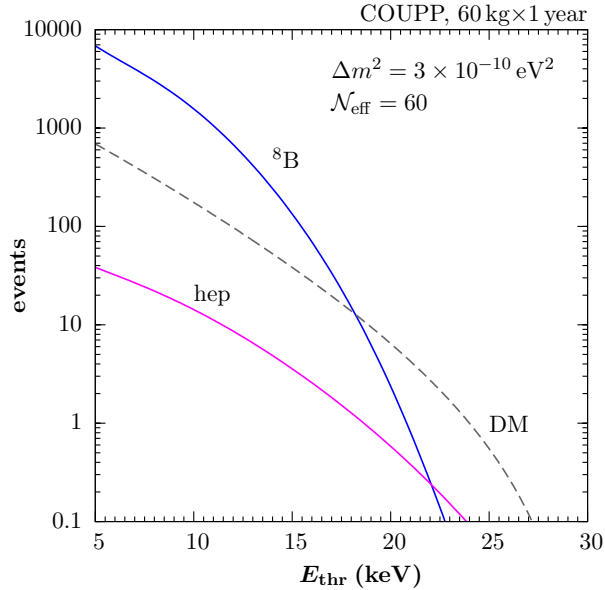} 
\caption{\small Predictions for a COUPP 60~kg bubble chamber for $\dmb
  = 3\times 10^{-10}\,\eV^2$ and $\Neff = 60$ as a function of bubble
  nucleation threshold with one year of exposure. The solid lines are
  from top to bottom for the \bet\ and hep fluxes of \nub. For
  comparison, a signal from a 10\,\GeV\ DM particle with
  spin-independent nucleon cross section of $\sigma_n =
  10^{-41}\,\cm^2$ is shown.}
\label{coupp}
\end{figure}

As already mentioned in Sec.~\ref{sec:simple}, detectors which employ
fluorocarbon compounds as target material are attractive because of
their favorable kinematics in contrast to heavier targets. A large
scale experiment of this type is the COUPP 60~kg bubble chamber
currently in the progress of moving into the SNOLAB underground
facility~\cite{Ramberg:2010zz}. It uses a superheated CF$_3$I liquid
with temperature and pressure adjusted such that only nuclear recoils
set off bubble nucleation events. It is a counting experiment for
events above adjustable threshold without \textit{a priori} insight
into the recoil energy distribution. We assume an exposure of 1~yr
together with a detector efficiency $\varepsilon=0.7$.

Light target nuclei make COUPP particularly attractive for the
searches of light WIMPs and \nub. In Fig.~\ref{coupp} the integral
signal of \nub\ for $\dmb = 3\times 10^{-10}\,\eV^2$ and $\Neff = 60$
as a function of detector threshold is shown. The two solid curves
from top to bottom correspond to \bet\ and hep neutrinos,
respectively. Already below a threshold energy of $\sim 20\,\keV$ the
\bet\ flux induces a clear signal. Limited insight into the energy
distribution should be possible by varying the rather ``steplike''
detector threshold. In particular, with a multi-year exposure (or
larger values of \Neff) the crossover from the \bet\ to the hep
neutrino spectrum may be observable. More importantly, the variation
of the threshold provides discriminating power between a putative DM
signal and \nub. The dashed line in Fig.~\ref{coupp} shows the
integral event rate for a 10\,\GeV\ DM particle with spin-independent
WIMP-nucleon cross section of $\sigma_n = 10^{-41}\,\cm^2$.

\section{Neutrino searches}
\label{sec:ns}

Here we are going to consider the elastic scattering of \nub\ from the
sun in solar neutrino experiments. The NC channels in those
experiments are not necessarily sensitive to this class of new
physics, given that the associated inelastic reactions exhibit the
scaling (\ref{eq:scaling}).  As shown in \cite{Pospelov:2011ha}, the
NCB interaction does not yield an observable rate for the D-breakup at
SNO. There is also the possibility of inelastic excitation of $^{12}$C
with subsequent emission of a 4.44\,\MeV\ $\gamma$. The analysis of
inelastic processes falls outside of the scope of the present study.

The experiments which are capable of detecting the elastic NCB signal
employ hydrocarbon scintillators, namely, Borexino and KamLAND as well
as the upcoming experiment SNO+. The dominant background at lowest
energies comes from $^{14}$C contamination of the mineral oil which
decays with a maximum $\beta $ energy of $Q = 156\,\keV$. Other, less
prominent backgrounds by long-lived isotopes at lowest energies are
the $\beta$-decays of $^{85}$Kr $(Q=687\,\keV)$ and $^{210}$Bi
$(Q=1.16\,\MeV)$.  Unfortunately, the decays from $^{14}$C prevent
sensitivity to \nub\ in current detectors. For example, the Borexino
experiment which uses dedicated mineral oil with low residual $^{14}$C
content measured its concentration to be
$^{14}\mathrm{C}/^{12}\mathrm{C}\simeq 2\times 10^{-18}$. Though this
is a seemingly small ratio, it translates into a rate of approximately
$\Gamma \simeq 6\times 10^4\, \mathrm{events}/\day/\ton$ below
0.2\,\MeV. Therefore, in the following we restrict our analysis to a
future possibility when the level of $^{14}$C is much reduced so that
one may hope to gain sensitivity to \nub.

The scattering of \bet\ neutrinos on protons can produce a recoil
energy of $\Er\lesssim 0.5\,\MeV$. However, the proton recoil is
quenched and the \nub\ signal may not show itself above the $^{14}$C
peak.  In organic scintillators Birk's law provides a phenomenological
description of the scintillation light yield per unit path
length~\cite{Birks:1964zz},
\begin{align}
\label{eq:birk}
  \frac{dL}{dx} = L_0 \frac{dE/dx}{1+k_B dE/dx} , 
\end{align}
where $k_B$ is Birk's constant and $dE/dx$ is the ion stopping power
in the material. The formula interpolates between the limiting cases
of low energy losses (no quenching), $dE/dx \ll k_B^{-1}$ with linear
dependence of the light output $dL/dx\sim L_0 dE/dx$ and high energy
losses (quenching), $dE/dx \gg k_B^{-1}$ for which saturation occurs,
$dL/dx\sim L_0/k_B$. Equation (\ref{eq:birk}) suggests the
(non-linear) relation between recoil and quenched energy,
\begin{align}
\label{eq:quench}
  \Ev = \int_0^{\Er}  \frac{dE}{1+k_B\, dE/dx} .
\end{align}

\begin{figure}
\centering
\includegraphics[width=0.5\textwidth]{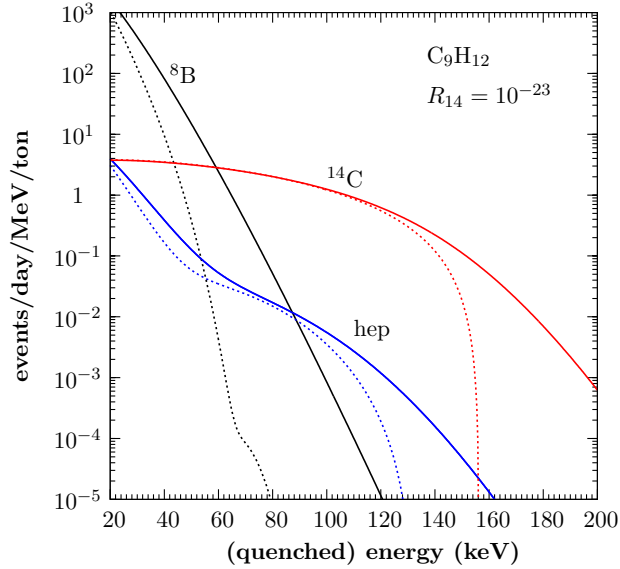} 
\caption{\small Liquid scintillator neutrino detector filled with
  pseudocumine, C$_9$H$_{12}$, and modeled after Borexino. Shown are
  the signals for \nub\ for $\Neff = 100$ and $\dmb = 2.5\times
  10^{-10}\,\eV^{2}$. The spectrum is largely dominated by background
  decays of $^{14}$C for which a contamination in carbon of one part
  in $10^{-23}$ has been assumed. Solid (dotted) curves are with
  (without) detector-resolution. Proton recoil energies are quenched;
  the quenching of electrons at lowest recoil energies has been
  neglected.}
\label{borexino}
\end{figure}

In the following we use Borexino, representative for other liquid
scintillator experiments. It has a a fiducial detector mass of
0.278\,kton filled with the scintillator pseudocumine, C$_9$H$_{12}$,
with a mass density of $\rho=0.88\,\mathrm{g}/\cm^3$.  We employ the
\texttt{SRIM} computer package to obtain the stopping power $dE/dx$ of
protons in pseudocumine. Quenched energies are then obtained from
(\ref{eq:quench}) with $k_B = 0.01\,\cm/\MeV$.  The scintillation
light yield is approximately $500\,\mathrm{PE}/\MeV$ and, for
simplicity, we assume a Gaussian energy resolution with $\sigma =
0.045\,\MeV\,\sqrt{\Ev}$~\cite{Dasgupta:2011wg} where $\Ev$ is in units
of~MeV.
 
We simulate the $^{14}$C background spectrum as follows. The decay
rate from $^{14}$C decay in C$_9$H$_{12}$ is given by
\begin{align}
  \frac{d\Gamma_{14}}{dE} = \frac{R_{14}}{t_{1/2}\ln{2}} \frac{9
    N_A}{M_{\mathrm{PC}}} \frac{df_{14}}{dE} = \frac{3.1\times
    10^4}{\day\times \ton} \left( \frac{R_{14}}{10^{-18}}
  \right)\times \frac{df_{14}}{dE}.
\end{align}
Here, $R_{14}$ is the ratio of $^{14}$C/$^{12}$C, $t_{1/2} = 5730\,$yr
is the half-life of $^{14}$C, $N_A$ is Avogadro's number, and
$M_{\mathrm{PC}} = 120.2~\mathrm{g/mol}$ is the molecular weight of
pseudocumine. The $\beta$-spectrum is given by~\cite{morita1973beta},
\begin{align}
\label{eq:c14spec}
  \frac{df_{14}}{dE} = \frac{1}{N}\times p_e E_e (E_0 - E_e)^2
  F(Z,E_e) C(E_e) ,
\end{align}
where $p_e$ and $E_e$ are the electron momentum and total energy,
$E_0$ is the total endpoint energy and $F(Z,E_e)$ is the Fermi
function for $^{14}$N; for the shape factor we use $C = 1 -
0.7\,E_e/\MeV$~\cite{Back:2002xz}. $N$ is chosen such that the
spectrum is normalized to unity. Electron recoils are somewhat
quenched by roughly $(10-40)\%$ for
$(100-10)\,\keV$~\cite{Back:2002xz}.  For simplicity we neglect this
complication here as well as more serious resolution effects near the
detector threshold. In a realistic scenario, the latter can affect the
spectrum below 100\,\keV\ substantially, see
\textit{e.g.}~\cite{Back:2002xz}.  Here we are merely concerned with
the question of how low the $^{14}$C content needs to be in order to
gain sensitivity to \nub\ recoils.

In Fig.~\ref{borexino} we compare the $^{14}$C background to a solar
\nub\ signal with parameters $\Neff = 100$ and $\dmb = 2.5\times
10^{-10}\,\eV^{2}$. The contamination has been fixed to $R_{14} =
10^{-23}$ which is already 5 orders of magnitude below the level for
the Borexino detector, $R_{14}(\mathrm{Borexino}) \simeq 2\times
10^{-18}$\cite{Alimonti:1998rc}. There is little hope that a signal
would show its presence above the $^{14}$C-decay endpoint energy of
$156\,\keV$. The detector resolution has the effect of smearing the
cutoff out to larger values of energy, hence burying the neutrino
signal. This can be seen by the difference between the dotted and
solid lines. At lowest recoil energies $E_R\lesssim 60\,\keV$ the
\bet\ signal dominates, but this may be a region which will not be
explorable in large mass scintillator experiments due to their
detector thresholds. Perhaps the only reasonable hope for using
carbon-based scintillators is the isotopic purification of not
excessively large, $O(100~{\rm kg})$, quantities of carbohydrates
with the eventual setup similar to the prototype of the Borexino
detector~\cite{Alimonti:1998rc}.

\section{Conclusions}
\label{sec:conclusions}

We have performed an extensive analysis of the model of \nub\
neutrinos that are sourced by the sun, and elastically scatter on
nuclei in underground dark matter experiments.  The goal of this study
was to assess the viability of this model as the explanation for the
reported anomalies, that are often interpreted as a possible dark
matter scattering signal.  Our findings are summarized in the master
plot of Figure 6, that is a direct analogue of the WIMP-mass
vs.~scattering cross section plot for DM scattering. We are now able
to conclude the following:

\begin{itemize}

\item {\em On the positive side for the model}, the \nub\ scenario
  with an oscillation length comparable to the earth-sun distance and
  with an effective enhancement of the NCB current by $O(100)$ is
  currently not seriously challenged by any of the existing
  experiments. On the contrary, this corner of parameter space
  provides a natural fit to the CoGeNT excess, and to the CRESST
  anomaly. Given enough uncertainty in the current status of the
  CoGeNT excess, and in the background-contaminated CRESST events, it
  is not difficult to see that the same regions of the parameter space
  of \nub\ model can be responsible for these anomalies. The null
  results of many experiments that challenge WIMP explanations of
  these anomalies (such as CDMS and Xenon100), do not challenge \nub\
  model.

\item {\em On the negative side for the model}, the values
  $\Neff\simeq 200$ which are necessary to fit the DAMA modulation
  amplitude are challenged by the CDMS-II low threshold analysis and
  by recent CRESST-II results.  It should be noted however, that in
  the presence of background~\cite{Kudryavtsev:2010zza}, smaller
  values for $\Neff$ become viable again. This requires larger
  modulation amplitudes of the signal, but this can presumably be
  achieved.  Also the phase of the predicted modulation signal is
  deviant from the DAMA data.  Even though one can achieve a phase
  reversal and have a maximum in the \nub\ scattering rate in early
  July, this is still far away from the DAMA phase, which corresponds
  to a modulation maximum in late May-early June. We do not know how
  to ``correct'' the model per se for this residual discrepancy.
  Further tension for the model may arise from the number of
  events predicted for the ``ionization-only'' signal at Xenon10. At
  face value, the parameter space of the model can potentially be
  constrained down to $\Neff\sim 40$, but the severity of constraint may
  well be mitigated by a poorly known energy calibration at those
  lowest recoils.

\item{\em Future prospects} for probing \nub-scattering look
  reasonably bright. In contrast to many DM models, the \nub\
  scattering pattern is fixed and we can make definite predictions as
  functions of only two parameters. For example, these predictions
  show that the COUPP experiment may be particularly sensitive to the
  \nub\ scattering signal. One can also conclude that a re-designed
  iteration of the CRESST-II experiment with reduced backgrounds will
  likely be very sensitive to the \nub\ model. Furthermore, a reduced
  electron-like background in the Xenon100 experiment will be able to
  probe the parameter space once the uncertain values for $Q_y$ and/or
  ${\cal L}_{\rm eff}$ are clarified.

\end{itemize} 

Finally, as this paper was readied for the submission, a new preprint
appeared~\cite{Harnik:2012ni} that examines a similar set of ideas.
It expands the set of interesting mediation mechanisms beyond NCB to
{\em e.g.} $B-L$ and ``massive photon'' forces with extra-light
mediators in the sub-MeV mass range. Unlike the \nub\ model, such
modifications may already be under strong tension from astrophysical
and cosmological constraints. We plan to return to \nub-related
signatures in astrophysical and cosmological settings in future work.

\bibliography{biblio}

\end{document}